\documentclass[onecolumn,superscriptaddress,pra]{revtex4}%
\usepackage{amsmath}
\usepackage{amsfonts}
\usepackage{amssymb}
\usepackage{graphicx}

\newcommand{\R}{\mathbb{R}}

\newcommand{\tr}{\mathrm{Tr}}
\newcommand{\Tr}{\mathrm{Tr}}

\newcommand{\cH}{\mathcal{H}}

\newcommand{\ra}{\rangle}

\newcommand{\ket}[1]{\left | \, #1 \right \rangle}
\newcommand{\bra}[1]{\left \langle #1 \, \right |}

\newcommand{\be}{\begin{equation}}
\newcommand{\ee}{\end{equation}}
\newcommand{\ba}{\begin{eqnarray}}
\newcommand{\ea}{\end{eqnarray}}

\usepackage{color}

\begin{document}

\title[Entanglement Typicality]{Entanglement Typicality}

\author{Oscar C. O. Dahlsten}
\address{Clarendon Laboratory, University of Oxford, Parks Road, Oxford OX1 3PU, UK}
\email{oscar.dahlsten@physics.ox.ac.uk }

\author{Cosmo Lupo}
\address{MIT, Research Laboratory of Electronics, 77 Massachusetts Avenue, Cambridge MA 02139, USA}
\email{clupo@mit.edu}

\author{Stefano Mancini}
\address{School of Science and Technology, University of Camerino, I-62032 Camerino, Italy}
\address{INFN-Sezione di Perugia, Via A. Pascoli, I-06123 Perugia, Italy}
\email{stefano.mancini@unicam.it}

\author{Alessio Serafini}
\address{Department of Physics \& Astronomy, University College London, 
Gower Street, London WC1E 6BT, UK}
\address{Scuola Normale Superiore, I-56126 Pisa, Italy}
\email{serale@theory.phys.ucl.ac.uk}

\begin{abstract}
We provide a summary of both seminal and recent results on typical entanglement. 
By ``typical'' values of entanglement, we refer here to values of entanglement quantifiers
that (given a reasonable measure on the manifold of states) appear with arbitrarily high 
probability for quantum systems of sufficiently high dimensionality.
We shall focus on pure states and work within the Haar measure framework for discrete quantum variables, where we report 
on results concerning the average von Neumann and linear entropies as well as arguments 
implying the typicality of such values in the asymptotic limit. We then proceed to discuss 
the generation of typical quantum states with random circuitry. Different phases of entanglement, 
and the connection between typical entanglement and thermodynamics are discussed. 
We also cover approaches to measures on the non-compact set of Gaussian states of continuous variable
quantum systems.
\end{abstract}

\maketitle 

\section{Introduction}

The term ``quantum entanglement'' was coined by Schr\"odinger \cite{Sch36} in connection with the criticism against quantum theory put forward by Einstein, Podolsky and Rosen \cite{EPR35}. Since then, it became synonymous with quantum correlations that cannot be explained by any local and real (hence classical) theory. As such it was traditionally relegated to foundational (and philosophical) issues until a few decades ago. 
In fact, in the nineties its usefulness in quantum information processing was realized, with the seminal protocols of superdense coding and teleportation \cite{BW92, Betal93}. Then it started to be considered as a resource, and consequently much attention has been devoted to its characterization and quantification through the introduction of suitable measures (see e.g. \cite{PV06, Hetal09}).

While the structure of two-particle entanglement has been almost thoroughly explored, it has become evident that the
extension to many-particles systems results in a prohibitive task \cite{PV06, Hetal09}. This is because 
complexity (and diversity) of multi-particle entanglement grows exponentially with the number of particles. 
A viable approach towards a characterization of entanglement in systems with many constituents, consists in focusing on the ``typical'' entanglement. 
Here by ``typical'' we mean the type of entanglement that appears with arbitrarily high probability in a quantum system
of sufficiently high dimensionality. 
This subsumes the use of random states (first introduced in \cite{L78}) so that an entanglement measure becomes a function of a random variable, hence a random variable itself, and will have an associated probability distribution. Then 
the strategy is aimed at simplifying the problem at hand by restricting the attention to only those states 
corresponding to the most pronounced part of the above probability distribution and neglecting the others. 
To sample random states, it is reasonable to resort to the most unbiased probability measure, that is the one emerging from the Haar measure of the unitary group (the invariant measure under application of any unitary transformation) whose elements allow one to get any state when applied to a given starting pure state. 
Of course the consideration of typical entanglement returns exact results 
if the distribution of the entanglement measure will become strongly peaked in the limit of a large number of particles.
In this paper we review the achievements concerning typical bipartite entanglement for random quantum states involving a large number of particles. We shall emphasize the statistical properties when such a number tends to infinity and discuss the finite size effects as well. 

Besides its interest in the context of quantum information theory, typical entanglement has also been put forward 
as a fundamental explanation for the emergence of thermodynamics, where randomised global quantum states result 
in mixed, `thermal' local Gibbs states. We shall also concisely review this area of application of entanglement typicality.

The layout of the paper is the following. Section \ref{entsec} recalls basic notions about pure states' entanglement and its measures. Then, random quantum states will be presented in Section \ref{randsec} and their generation 
through random quantum circuits discussed in Section \ref{circuitsec}.
The statistical properties of typical entanglement in terms of different phases are detailed in Section \ref{statsec} and the relations to thermodynamics considered in Section \ref{thermosec}. Finally, in Section \ref{cvsec} the issue of typical entanglement will be addressed within the continuous variable framework. Conclusions are drawn in Section \ref{conclusec}. \ref{appendix} contains a cursory treatment of random mixed states of many qubits.


\section{Entanglement in a nutshell}\label{entsec}

We summarize here basic notions about the entanglement of pure states and refer the reader to recent reviews 
on the subject of quantum entanglement for further details \cite{Hetal09,PV06}. 

Let us consider a bipartite system with associated Hilbert space $\cH_A\otimes\cH_B$.
Then, any bipartite pure state $|\Psi_{AB}\ra\in\cH_A\otimes\cH_B$ is said to be \emph{separable}
if there exist  $\ket{\psi_A}\in\cH_A, \ket{\psi_B}\in\cH_B$ such that  
\be
\ket{\Psi_{AB}}=\ket{\psi_A}\ket{\psi_B},
\ee 
i.e. it can be written as a tensor product of vectors belonging to the Hilbert spaces of the subsystems\footnote{For pure states the notion of separability coincides with that of factorability.}.
On the contrary, if there not exist any $\ket{\psi_A}\in\cH_A, \ket{\psi_B}\in\cH_B$ of such kind,
the state $|\Psi_{AB}\ra$ is said to be \emph{entangled}.

Quite generally, given orthonormal bases $\{\ket{e_A^i}\}_i$ for $\cH_A$ and $\{\ket{e_B^j}\}_j$ for $\cH_B$, 
we can write 
\be
\label{PsiAB}
\ket{\psi_{AB}}=\sum_{i=1}^{N_a}\sum_{j=1}^{N_B}\Psi_{ij} \ket{e_A^i}\ket{e_B^j},
\ee 
where $N_A=\dim\cH_A$ and $N_B=\dim\cH_B$. Then it turns out that
the state $\ket{\psi_{AB}}$ is separable if and only if the matrix $\Psi$ of coefficients $\Psi_{ij}$ has rank one.
Furthermore, there exists a bi-orthonormal basis $\{\ket{\tilde{e}_A^i}\ket{\tilde{e}_B^i} \}_{i}$ for $\cH_A\otimes\cH_B$
where $\ket{\psi_{AB}}$ takes the form
\be
\ket{\psi_{AB}}=\sum_{i=1}^{\min[N_A,N_B]} \lambda_i \ket{\tilde{e}_A^i}\ket{\tilde{e}_B^i},
\ee 
known as Schmidt decomposition. The quantities $\lambda_i$ (Schmidt coefficients) are non-zero singular 
eigenvalues of $\Psi$, or in other words $p_i = \lambda_i^2$ are non-zero eigenvalues of either reduced 
density operator $\rho_A=\tr_B(\ket{\psi_{AB}}\bra{\psi_{AB}})$ or $\rho_B=\tr_A(\ket{\psi_{AB}}\bra{\psi_{AB}})$. 

Entanglement is invariant under local unitary operations $U_A\otimes U_B$. 
Since the coefficients $\lambda_i$ (or equivalently the coefficients $p_i$) are the only parameters invariant under such transformations, 
they completely determine the bipartite entanglement.

A quantitative measure of entanglement must satisfy two fundamental properties\footnote{It is still not clear if the convexity is a further mandatory ingredient.}:
\begin{itemize}
\item[i)]
It cannot increase under local operation and classical communication;
\item[ii)]
It must vanish for separable states;
\end{itemize}
In addition we may require normalisation so that the amount of entanglement is $\log N$ when 
$\ket{\psi_{AB}}=\sum_{i=1}^N  \ket{\tilde{e}_A^i}\ket{\tilde{e}_B^i}/\sqrt{N}$, i.e. for a maximally entangled state.

The \emph{entropy of entanglement}, i.e. entropy of subsystem (either $A$ or $B$)
\be
\label{ententdef}
{\cal S}_A=-\tr\rho_A\log\rho_A={\cal S}_B=-\tr\rho_B\log\rho_B,
\ee
satisfies these conditions and will be considered throughout this paper\footnote{In the asymptotic regime all pure states entanglement measures correspond to this one.}.
More generally, one could consider quantum Renyi entropy replacing the von Neumann entropy, that is 
\be
{\cal S}_A^q(\rho_A)=\frac{1}{1-q}\log\tr(\rho_A^q).
\ee
For $q=1$ it reduces to (\ref{ententdef}), while for $q=2$ 
it is related to the so called \emph{purity}
\be
\label{puritydef}
{\cal P}_A=\tr\left(\rho_A^2\right) 
\ee
by the relationship
\be
{\cal S}_A^2(\rho_A)=-\log{\cal P}_A \, .
\ee

\subsection{Separability and entanglement of continuous variables}\label{CVsubsec}

The extension of the above arguments to infinite dimensional Hilbert spaces presents some oddities.
Systems with associated Hilbert space isomorphic to $\ell^2(\mathbb{C})$ or $L^2(\mathbb{R})$ are 
usually  referred to as continuous variable (CV) systems.
 
For a bipartite CV pure state living in  $\ell^2(\mathbb{C})_A\otimes \ell^2(\mathbb{C})_B$ (where Fock bases are standardly used) 
separability occurs if and only if it has Schmidt 
rank equal to one, i.e. one Schmidt coefficient equal to one and all the other zero. However there cannot be maximally entangled states simply because a state with all 
Schmidt coefficients equal would lead to an infinite norm.
Moreover, the set of separable (mixed) states is nowhere dense \cite{CH00} and as consequence would have volume zero, in contrast to what happens in finite dimension \cite{Zetal98}. 
 
The entropy of entanglement remains a good measure for pure CV states. However, it is unbounded and becomes infinite for certain states. 
To avoid this problem one has to impose suitable constraints.
That can be more easily done within the set of Gaussian states (see \cite{BV05,Wetal12} for reviews on Gaussian states in quantum information). 

A Gaussian states $\rho_{AB}$ of $2$ modes on the Hilbert space $L^2(\mathbb{R})_A\otimes L^2(\mathbb{R})_B$ of functions of 
position variables $(q_A,q_B)$, is characterized by the displacement vector (first moments)
\be
 d_i=\tr(\rho_{AB} R_i),
 \ee
 and the covariance matrix (second moments) 
 \be
 \sigma_{ij}=\tr(\rho_{AB}\{R_i-d_i,R_j-d_j\}),
 \ee 
 where 
$R=(Q_A,P_A,Q_B,P_B)$ and $Q$, $P$ are mode position and momentum observables respectively.
These operators satisfy the canonical commutation relations $[R_k,R_{k'}]= 2 i J_{kk'}$ where 
\be
J=\bigoplus_{i=1}^2 
\left(
\begin{array}{cc}
0 & 1 \\
-1 & 0
\end{array}
\right),
\ee
is the symplectic form (we assume the convention that the covariance matrix of the vacuum state is set to $I$).

Since the displacement $d$ can be removed by local unitary operations, only the covariance matrix $\sigma$
is relevant for entanglement. 
Notice that Heisenberg uncertainty imposes that \cite{BV05,Wetal12}
\be\label{Hrel}
\sigma+i J\ge 0.
\ee
The covariance matrix $\sigma$ describes a pure state if and only if $(\sigma J)^2=-I$.

In order to avoid divergences of physical quantities it is standard within the manifold of Gaussian states to constrain the mean value of the `energy' (assuming free, non interacting oscillators) in each subsystem. This amounts to fixing the following values
\ba
E_A&=&\Tr\{\rho_{AB}(a^\dag a+a a^\dag)\}, \label{Vener}\\
E_B&=&\Tr\{\rho_{AB}(b^\dag b+b b^\dag)\},
\ea
where $a,a^\dag$ (resp. $b,b^\dag$) are ladder operators, with real and hermitian part given by
$Q_A,P_A$ (resp. $Q_B,P_B$), that provide a natural link to $\ell^2(\mathbb{C})_A\otimes \ell^2(\mathbb{C})_B$.  

Now, suppose that $\sigma$ describes a bipartite Gaussian pure state $\rho_{AB}$ of $1 + 1$ modes, 
then the reduced density operator $\rho_A$ will be still Gaussian and characterized by a covariance matrix $\sigma_A$. 
The latter can always be diagonalized by means of some symplectic matrix $S_A$, so to have 
$S\sigma_A S^T=diag(\nu,\nu)$ with $\nu \in [1,\infty)$ is the so called symplectic eigenvalue of $\sigma_A$
(we have that the symplectic eigenvalues of $\sigma_A$ equals those of $\sigma_B$ if $\rho_{AB}$ is a pure Gaussian state).
Similarly, for a bipartite Gaussian state of $n_A + n_B$ the symplectic diagonalization defines a set of
$n_A$ (assuming $n_A \leq n_B$) symplectic eigenvalues, $\nu_1, \nu_2, \dots, \nu_{n_A}$. 
One can show that any entanglement measure is a function of the symplectic eigenvalues only.
In particular, the entropy of entanglement (\ref{ententdef}) reads \cite{HW01}
\be
{\cal S}_A=\sum_{i=1}^{n_A} h(\nu_i),
\ee
where 
\be
h(x)=\frac{x+1}{2}\log\left(\frac{x+1}{2}\right)
-\frac{x-1}{2}\log\left(\frac{x-1}{2}\right).
\ee
The purity (\ref{puritydef}) can be expressed in terms of symplectic eigenvalues too, as
\be
{\cal P}_A=\left[\prod_{i=1}^{n_A} \nu_i\right]^{-1}.
\ee


\section{Random quantum states}\label{randsec}

The set of pure states on a $N$ dimensional Hilbert space $\cH$ forms a complex projective space $\mathbb{C}P^{N-1}$ on which there exists a natural uniform measure in the sense that, because of its unitary invariance, it equally weighs different regions of the space. This measure is constructed by borrowing the Haar measure on $U(N)$. In this framework, to generate a random pure state one applies a random  
unitary $U\in {\rm U}(N)$ to a fixed state $\ket{\psi_0}\in\cH$, which is equivalent to taking a vector (column) from a random unitary $U\in {\rm U}(N)$. This particular choice of a measure on the set of quantum states enjoys the 
privilege of being invariant under any Hamiltonian evolution which, in a sense, establishes 
a connection with dynamics.

Being ${\rm U}(N)$ isomorphic to a $N^2$ dimensional manifold embedded on $\R^{2N^2}$,
expressing the Haar measure there implies the use of $N^2$ local coordinates. 
Exploiting the Hurwitz parametrization, that generalizes the Euler angles, these read \cite{B05} 
\be 
0\le \alpha < 2\pi,\quad
0\le \vartheta_{k\ell}\le \frac{\pi}{2},\quad
0\le \varphi_{k\ell} < 2\pi,\quad
0\le \chi_{\ell} < 2\pi,
\ee
with $1\le k < \ell \le N$. 
Then the measure on ${\rm U}(N)$ (normalized to 1) can be written as \cite{ZK94}
\ba
\label{dmuU}
d\mu(U)=\frac{[N(N-1)]!}{2^N\pi^{N^2}} d\alpha \prod_{1\le k < \ell \le N} \cos\vartheta_{k\ell}\left(\sin\vartheta_{k\ell}\right)^{2k-1} 
d\vartheta_{k\ell} \, d\varphi_{k\ell} \prod_{1<\ell\le N} d\chi_{\ell}.\nonumber\\
\ea
Notice that, being ${\rm U}(N)={\rm U}(1)\times {\rm SU}(N)$ (with $\alpha$ the ${\rm U}(1)$ parameter), from (\ref{dmuU}) we can also get the measure on ${\rm SU}(N)$.
In turn, the measure on $\mathbb{C}P^{N-1}$ can be derived by observing that $\mathbb{C}P^{N-1}={\rm SU}(N)/{\rm U}(N-1)$ and it results (normalized to 1) \cite{ZS01} 
\be
\label{dmuCP}
d\mu(\ket{\psi})=\frac{(N-1)!}{\pi^{N-1}}\prod_{1\le k\le N-1} \cos\vartheta_k(\sin\vartheta_k)^{2k-1} d\vartheta_k d\varphi_k,
\ee
where
\be 
0\le \vartheta_{k}\le \frac{\pi}{2},\quad
0\le \varphi_{k} < 2\pi,
\ee
with $1\le k  \le N-1$.

Let us now discuss the induced measure on the set of reduced density operators for a bipartite system.
Consider a bipartite quantum system with Hilbert space $\cH_A\otimes\cH_B$ of dimension $N_A\times N_B$.
A pure state $\ket{\psi_{AB}}\in\cH_A\otimes\cH_B$ can be expanded as in Eq.(\ref{PsiAB}) 
and thus can be represented by a rectangular ($N_A\times N_B$) matrix $\Psi_{ij}$. 
Upon normalization, the ensemble of uniformly distributed pure states coincides with the Ginibre
ensemble of random matrices with i.i.d.\ Gaussian distributed entries with zero mean and finite variance. 
In turn, the density operator 
$\ket{\psi_{AB}}\bra{\psi_{AB}}$ will be represented by a square ($N_AN_B\times N_AN_B$) matrix $\Psi_{ij}\Psi_{i'j'}^*$. 
The partial trace with respect to the subspace $\cH_B$ gives the reduced density matrix (of the $A$ subsystem)
\be
\label{rhopsipsi+}
\rho_{ij}=\sum_k \Psi_{ik}\Psi_{jk}^*, \quad i,j=1,\ldots,N_A,
\ee 
where we have omitted the label indicating the $A$ subsystem.
 
Assuming $N_B\ge N_A$, we write as consequence of (\ref{rhopsipsi+})  $\rho=\Psi\Psi^\dag$. Then, we have the following distribution of (hermitian) matrices
\be
P(\rho)\propto \int d\mu\left(\Psi\right) \delta\left(\rho-\Psi\Psi^\dag\right)\delta\left(\tr\Psi\Psi^\dag-1\right),
\ee
where the measure $d\mu\left(\Psi\right)\equiv d\mu(\ket{\psi})$ is the one in $\mathbb{C}P^{N_AN_B-1}$ (see (\ref{dmuCP})).
Furthermore, the first delta imposes that $\rho=\Psi\Psi^\dag$, while the second one imposes the unit trace.
That is, the distribution of the reduced density matrix coincides (upon normalization) to the 
distribution of the Wishart matrix $\Psi\Psi^\dag$.

Making the change of variable $\Psi=\sqrt{\rho}\tilde\Psi$, $d\mu\left(\Psi\right)=\det\rho^{N_B}d\mu\left(\tilde\Psi\right)$ and noticing that 
$\delta\left(\sqrt{\rho}\left(1-\tilde{\Psi}\tilde{\Psi}^\dag\right)\sqrt{\rho}\right)=\det\rho^{-N_A}\delta\left(1-\tilde{\Psi}\tilde{\Psi}^\dag\right)$
we obtain
\be
P(\rho)\propto\theta(\rho)\delta(\tr\rho-1)\det\rho^{N_B-N_A},
\ee
where the theta function guarantees the positivity of $\rho$. Now, being $\rho$ unitarily diagonalizable we write it 
as $\rho=U\Lambda U^\dag$ with $\Lambda=diag\{p_1,\ldots,p_{N_A}\}$.
Then, by integrating over $d\mu(U)$ given by (\ref{dmuU}) we obtain the joint eigenvalues density distribution
(this expression was first derived in \cite{LP88})
\ba
\label{Pla}
&&P(p_1,\ldots,p_{N_A})=\int d\mu(U) P(U\Lambda U)\nonumber\\
&&= Z^{-1}  \delta\left(1-\sum_i p_i\right)
\prod_{i<j}(p_i-p_j)^2\prod_i p_i^{N_B-N_A}\theta(p_i), 
\ea
where the normalization constant reads \cite{ZS01}
\be
Z = \frac{\prod_{j=0}^{N_A-1}\Gamma(N_B-j)\Gamma(N_A-j+1)}{\Gamma(N_AN_B)},
\ee
with $\Gamma$ denoting the Euler Gamma function.

At this point we may notice that (\ref{Pla}) is the distribution of the squared 
Schmidt coefficients of the state $\ket{\psi_{AB}}$. 
By using (\ref{Pla}) one can derive the distribution for the subsystem entropy ${\cal S}$
\be
P({\cal S})=\int dp_1 \cdots dp_{N_A} \, P(p_1,\ldots,p_{N_A}) \delta\left({\cal S}+\sum_i p_i \log{p_i}\right).
\label{Sdistribution}
\ee
which results in the mean value
\be
\label{Save}
\mathbb{E}\left({\cal S}(\rho)\right)=\frac{1}{\ln 2}\left(\sum_{k=N_B+1}^{N_AN_B} \frac{1}{k}-\frac{N_A-1}{2N_B}\right).
\ee
This expression was first conjectured by Page \cite{P93} and then proved later on by several authors 
\cite{Pageproof1,Pageproof2,Pageproof3}. 

Recall that the subsystem entropy is an entanglement measure for bipartite pure states and its maximum is 
$\log N_A$. From (\ref{Save}) it follows \cite{L78}
\be
\label{Savebound}
\mathbb{E}\left({\cal S}(\rho)\right)>\log N_A -\frac{N_A}{N_B\ln 2},
\ee
which scales like $\log N_A$ when $N_A\simeq N_B$.

Likewise we can derive the expectation value of any Renyi entropy.
Then, for $q=2$, the expectation value of the purity can be derived:
\be
\mathbb{E}\left({\cal P}(\rho)\right)=\frac{N_A+N_B}{N_AN_B+1}.
\label{Pave}
\ee

Now a value of random variable, like ${\cal S}(\rho)$, is called \emph{typical} if its probability distribution is peaked around the mean.
Actually, in Ref.\cite{HLW06} it has been shown that the probability that a state $\ket{\psi_{AB}}$ drawn randomly from $\cH_A\otimes \cH_B$ gives the subsystem entropy smaller than $\log N_A$ is exponentially smaller; more precisely 
\be 
{\rm Pr}\left\{
{\cal S}(\rho_A)<\log N_A-\frac{N_A}{N_B\ln 2}-\alpha \right\} \le \exp\left[-\frac{(N_A N_B-1)}{8\pi^2\ln 2(\log N_A)^2}\alpha^2\right].
\ee
This is referred to as concentration measure effect and comes form the concentration of the spectrum of the reduced density matrix of a bipartite system when the dimensions of both subsystems become large. This effect can be traced back to the fact that the uniform measure on the $k$-sphere $\mathbb{S}^k$ concentrates about any equator as $k$ gets large and any polar cap smaller than the hemisphere has a relative volume exponentially smaller in $k$. This implies that similar results hold true for the value of any smooth function on the sphere: all such functions will be overwhelmingly likely to take values close to the average except for a set of volume exponentially small in $k$.
Hence, random pure states are typically highly entangled (though not necessarily maximally entangled). 

The complete distribution of bipartite entanglement of random pure states may be exactly reconstructed in terms of purity \cite{G07,Fetal08}, von Neumann entropy \cite{Florio,kumar,NMV11}, and all Renyi entropies with $q > 1$ \cite{NMV10}. 
For a possible approach to typical multipartite entanglement, see \cite{fac1,fac2}.

The extension of the above arguments from the case of pure states $\ket{\psi_{AB}}$ to the case of mixed states 
$\rho_{AB}$ is briefly discussed in \ref{appendix}.


\section{Random quantum circuits}\label{circuitsec}

We provide here an operational interpretation of typical entanglement by means of random circuits.

Given a set of $n$ qubits, 
a random circuit $C_\ell$ is a product $W_\ell \ldots W_1$ of two-qubit gates where each $W_i$ is independently constructed in the following way: a pair of distinct integers $c\neq t$ is randomly and uniformly chosen from ${1, \ldots, n }$. Then, single-qubit unitaries $U[c]$ and $V[t]$ acting on qubit $c$ and $t$ respectively are drawn independently from the uniform measure on ${\rm U}(2)$. Finally, $W = {\rm CNOT}[c,t]U[c]V[t]$ where ${\rm CNOT}[c,t]$ is the controlled-NOT gate with control and target qubit $c$ and $t$ respectively.

Since the universal set (of all one qubit gates together with CNOT) 
can generate the whole of ${\rm U}(2^n)$, such random circuits can produce any unitary. 
This process converges to a unitarily invariant distribution, but the Haar distribution is unique, hence the resulting unitary will be uniformly distributed in ${\rm U}(2^n)$. 
However, the convergence rate results exponentially slow in the number of qubit $n$, since approximating an arbitrary unitary to a given accuracy using a set of fixed size gates requires a number of steps that grows exponentially with 
$n$ \cite{K95}. 
Thus obtaining the uniform distribution to a fixed accuracy may look unphysical.

On the other hand, entanglement gives rise to peculiar properties of a quantum state. Then, their faithful reproduction may be possible with fewer physical resources, i.e. elementary gates, than those required for the generation of the expectation value for an arbitrary observable.
Ref. \cite{ODP07} explored whether typical entanglement properties can be obtained efficiently, i.e. polynomially in the number of qubits, using only one- and two-qubit gates. 

There, it was considered the set of $n$-qubits split into two subsets $A$ (with $n_A$ qubits) and $B$ (with $n_B$ qubits). Let  $\ket{\psi_0}$ be a initial state in $AB$ and consider a random circuit $C_\ell$ consisting of $\ell$ randomly chosen two-qubit quantum gates. Defining $\ket{\psi_\ell} = C_\ell \ket{\psi_0}$ the amount of entanglement 
of reduced density operator $\rho_{A,\ell} = \tr_B (\ket{\psi_\ell}\bra{\psi_\ell})$ of subsystem $A$ will be ${\cal S}(\rho_{A,\ell})$ according to Eq.(\ref{ententdef}).
Then, in Ref. \cite{ODP07} it was shown that, independently of the initial state $\ket{\psi_0}$, convergence of the expected entanglement to its 
asymptotic value to an arbitrary fixed accuracy $\epsilon$ is achieved after a number of random two-qubit gates that is polynomial in the number of qubits. 
More precisely, given $n_B \ge n_A$, $\epsilon \in (0, 1)$ and a number $\ell$ of gates in $C_\ell$ satisfying
\be
\ell \ge 9n (n - 1) \frac{(3 \ln 2) n + \ln \epsilon^{-1} }{4},
\ee
we have 
\be
\mathbb{E}[{\cal S}(\rho_{A,n})]\ge n_A-\frac{(2^{n_A-n_B}+\epsilon)}{\ln2},
\ee
which is similar to the bound of Eq.(\ref{Savebound}).
The convergence occurs in approximately $n \log n$ steps, so the bound is not tight.

This can be explained as follows.
Writing $\ket{\psi_\ell}\bra{\psi_\ell}=2^{-n/2}\sum_{s\in\{0,x,y,z\}^n}\xi_\ell(s)\otimes_{i=1}^\ell\sigma^{s_i}[i] $,
where $\xi_\ell(s)=2^{-n/2}\tr\left(\otimes_{i=1}^n\sigma^{s_i}[i] \ket{\psi_\ell}\bra{\psi_\ell}\right)$ and $\sigma^{s_i}[i]$ is the $s_i$-th Pauli operator acting on the $i$-th qubit,
the reduced density operator $\rho_A={\tr}_{B}\left(\ket{\psi_\ell}\bra{\psi_\ell}\right)$ yields
\be
\mathbb{E}\left[\tr\left(\rho_A^2\right)\right]=2^{n_B}\sum_{s | s_i=0, \forall i\neq A}
\mathbb{E}\left[\xi_\ell^2( p)\right].
\ee
The coefficients $\mathbb{E}\left[\xi_\ell^2( p)\right]$ form a probability distribution on $\{0,x,y,z\}^n$ for all $\ell$, and these probabilities evolve as a Markov chain with transition matrix taking $p$ distributed according to 
$\left(\mathbb{E}\left[\xi_\ell^2(s)\right]\right)_p$ in one step to $q$ distributed according to $\left(\mathbb{E}\left[\xi_{\ell+1}^2(s)\right]\right)_q$.
Then, after a certain number of steps taken in the Markov chain, an abrupt approach to the stationary distribution occurs
giving rise to a  cut-off effect in the entanglement probability distribution \cite{ODP07}.
Ref.\cite{Detal07} discussed how to identify the transition from the phase of rapid spread of entanglement to the stationary phase where entanglement is typically maximal.

Actually, the efficient generation of typical entanglement features can be traced back to the fact that 
random circuits of only polynomial length form approximate 1- and 2-designs \cite{Dankert06, HL09}.
The notion of  $k$-designs quantifies the extent to which pseudo-random operators behave like the uniform distribution. Let $\{p_i,U_i\}$ be an ensemble of unitary operators and define 
\be
{\cal G}_W(\rho)=\sum_i p_i U_i^{\otimes k}\rho (U_i^\dag)^{\otimes k}, \quad
{\cal G}_H(\rho)=\int d\mu(U) U^{\otimes k}\rho (U^\dag)^{\otimes k}, 
\ee
then the ensemble is a unitary $k$-design if ${\cal G}_W={\cal G}_H$
and  is $\epsilon$-approximate unitary $k$-design if $\|{\cal G}_W-{\cal G}_H\|_{\diamond}\le \epsilon$, with $\|\bullet\|_\diamond$ the diamond norm \cite{K97}.

In Ref.  \cite{HL09} it is conjectured, based on an analogous classical result, that a random circuit on $n$ qubits of length $poly(n, k)$ is an approximate $k$-design. 
While this is not proved, it is instead rigorously showed that a circuit of length $O(n(n + \log 1/\epsilon))$ yields an 
$\epsilon$-approximate 2-design (so the first two moments are equal within $\epsilon$ to those of the Haar distribution). 
More recent results show that certain random circuits are actually approximate polynomial designs for any $k$ ~\cite{BrandaoHH12}. (This does not mean that the circuits approximately generate the Haar measure in a reasonable number of gates, since the statement is that if you fix $k$, then one needs only $poly(n)$ gates; it is not a statement about how the number of gates required scale with $k$ for fixed $n$.)

In Ref. \cite{PDP08} a different scheme to generate random quantum circuits that does not make use of classical random numbers was proposed. It relies on a particular type of entangled state called weighted 
graph state. 
Consider $n\times m$ qubits sitting on vertices of a simple graph $G$ embedded in a 2-dimensional lattice, each one in the state $\ket{+}$. Then a weighted graph state is
\be
\ket{\Psi_{WGS}}=\prod_{\{a,b\}\in E} U_{a,b}(A_{a.b}) \ket{+}^{\otimes n m} 
\ee
where the product is taken overall edges and unitaries are defined as
\be
U_{ab}=\exp\left[-i A_{a,b}\frac{\pi}{4}(I-\sigma_z^a)\otimes (I-\sigma_z^b)\right]
\ee
where $A_{a,b}$ are the entries of the adjacency matrix.
Suppose that the first column of quits represent the input state, 
then measurements are performed successively on columns $1$ through $m-1$ leaving the output state on the last column (the $m$-th).  Furthermore, on each column just performs projective measurements in the $\{ \ket{+}, \ket{-} \}^{\otimes n}$ basis. In this way the randomness of the measurement outcomes chooses the particular circuit and no classical random numbers are necessary. 

Moreover non-universal set of gates can also generate typically maximal entanglement. Stabilizer states, an important discrete subset of general quantum states on finite dimensions, have typically maximal entanglement~\cite{SmithL06, DahlstenP06} and applying gates that are universal for stabilizer states generates this typical entanglement in finite time~\cite{Detal07}. In addition it has been shown that random circuits of elementary diagonal (in the computational basis) unitary gates are also 2-designs, for a suitable definition of diagonal 2-designs \cite{NakataTM12,NakataM13}. The typical entanglement generated by these depends on the initial state, e.g. a product state $\ket{0}^\otimes n$ is left invariant but certain superposition product states will typically become maximally entangled. 


\section{Phase transitions of entanglement}\label{statsec}

We have seen in Sec. \ref{randsec} that uniformily distributed states are typically close 
to be maximally entangled. That is, the average entanglement (as quantified,
e.g., by the Renyi entropy of entanglement) is close to its maximum value, and
the probability of deviation from the average are exponentially suppressed in
the dimension of the quantum system.

Suppose to have a bipartite system of dimension $N_A \times N_B$, with $N_A \leq N_B$.
In terms of the squared Schmidt coefficients the typicality of entanglement is expressed by
the fact that for typical states we have $p_i \sim 1/N_A$.
However, this information alone is not sufficient to characterize the typical
distribution of the squared Schmidt coefficients (also known as {\it entanglement spectrum}).
This goal can be achieved by applying the method of stationary phase in the limit $N_A \to \infty$.
Moreover, the same method allows one to derive the explicit distribution of certain entanglement measures,
e.g., the Renyi entropies of order $q>1$.

Let us consider the integral of the probability density of the squared Schmidt coefficients given by Eq.(\ref{Pla}) 
\begin{eqnarray}
Z = \int_{p_i \geq 0} dp \, \prod_{i<j}(p_i-p_j)^2 \prod_i p_i^{N_B-N_A} \delta\left(\sum_i p_i-1\right).
\end{eqnarray}
It results
\begin{eqnarray}
Z  =  \int_{C(1)} dp \, \exp{\left[ (N_B-N_A)\sum_i\ln{p_i}+2\sum_{i<j}\ln{|p_i-p_j|} \right]},
\end{eqnarray}
where the integration is over the region $C(1)$ determined by the constraints $\sum_i p_i=1$ and $p_i \geq 0$.
The latter expression for $Z$ shows the formal analogy between the statistics of random states 
of a quantum system and the thermodynamics of a $2$-dimensional Coulomb gas (a well known fact
in random matrix theory).
According to this analogy, $Z$ corresponds to the partition function of $N_A$ charged particles
on a line with coordinates $p_i \in [0,1]$
(that interacts through the $2$-dimensional Coulomb potential $V = -2\sum_{i<j}\ln{|p_i-p_j|}$),
in the external potential $V_\mathrm{ext} = -(N_B-N_A)\sum_i\ln{p_i}$.

For $N_A \gg 1$ one can apply the method of stationary phase and evaluate
\begin{equation}
Z \simeq e^{-E_s},
\end{equation}
where 
\begin{equation}
E_s = \min_{p} \left[ - (N_B-N_A)\sum_i\ln{p_i} - 2\sum_{i<j}\ln{|p_i-p_j|} + \mu \sum_i p_i \right],
\end{equation}
is the minimum energy under the constraints $\sum_i p_1 = 1$ and $p_i \geq 0$, and
$\mu$ is the associated Lagrange multiplier.
Notice that the typical distribution of the squared Schmidt coefficients is the one
that minimizes the energy and is the solution of the stationary phase equation
\begin{equation}
\sum_{j|j\neq i} \frac{1}{p_j-p_i} = \frac{(N_B-N_A)}{2p_i} - \frac{\mu}{2} \, .
\end{equation}

In the limit $N_A \to \infty$ we replace the sum with an integral and obtain the equation
\begin{equation}
\mathsf{P.V.} \hspace{-0.1cm} \int_0^1 dp' \, \frac{\omega(p')}{p'-p} = \frac{N_B - N_A}{2p} - \frac{\mu}{2} \, ,
\end{equation}
which has to be solved under the constraint
\begin{equation}
\int_0^1 dp \, \omega(p) p = 1 \, ,
\end{equation}
where $\omega(p) = \sum_i \delta(p-p_i)$ is the density of the squared Schmidt coefficients
(satisfying $\int_0^1 dp \, \omega(p)= N_A$).
The solution is given by the Marchenko-Pastur distribution \cite{P93,MP} which, under the 
assumption of $N_A\le N_B$ made, reads
\begin{eqnarray}
\omega_s(p) = \left\{ 
\begin{array}{lr}
\frac{N_AN_B}{2\pi} \frac{\sqrt{(p-a)(b-p)}}{ p} & \mbox{for} \, \, p \in [a,b]  \\
0 & \mbox{for} \, \, p \not\in [a,b]  
\end{array}\right. ,
\end{eqnarray}
where
\begin{equation}
a = \left(\frac{1}{\sqrt{N_A}} -\frac{1}{\sqrt{N_B}} \right)^2 , \quad
b =   \left(\frac{1}{\sqrt{N_A}} +\frac{1}{\sqrt{N_B}} \right)^2 .
\end{equation}


Given an entanglement measure ${\cal E}(p)$, the same approach can be applied 
to compute its probability density
\begin{eqnarray}
P({\cal E}) & = & Z^{-1} \hspace{-0.2cm} \int_{p_i\ge 0} dp \, \prod_{i<j}(p_i-p_j)^2 \prod_i p_i^{N_B-N_A} \delta\left(\sum_i p_i-1\right) \delta \left( {\cal E}(p) - {\cal E} \right)\nonumber \\
& = & Z^{-1} \int_{C(1,{\cal E})} \hspace{-0.2cm} dp \, \exp{\left[ (N_B-N_A)\sum_i\ln{p_i}+2\sum_{i<j}\ln{|p_i-p_j|} \right]} \, ,
\end{eqnarray}
where the integration is over the region $C(1,{\cal E})$ determined by
{the constraints $\sum_i p_i = 1$, ${\cal E}(p) = {\cal E}$ and $p_i \geq 0$}.
For $N_A \to \infty$ the stationary phase approximation yields
\begin{equation}
P({\cal E}) \simeq Z^{-1} e^{-E_s({\cal E})} \, ,
\end{equation}
where $E_s({\cal E})$ is the minimum energy under the above constraints.
The corresponding solution $\omega_s(p|{\cal E})$ describes the entanglement spectrum 
of states belonging to the submanifold with ${\cal E}(p) = {\cal E}$.

\begin{figure}[ht]
\centering
\includegraphics[width=0.3\textwidth]{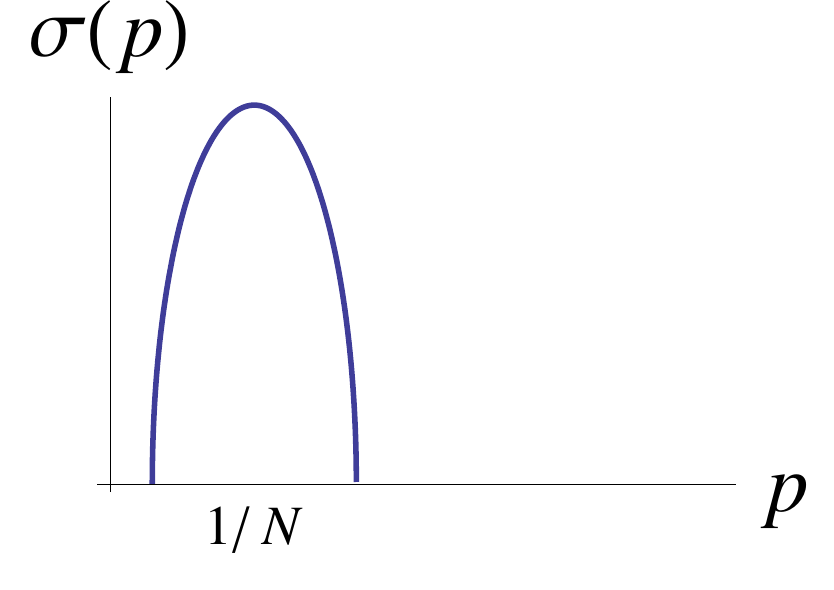}
\includegraphics[width=0.3\textwidth]{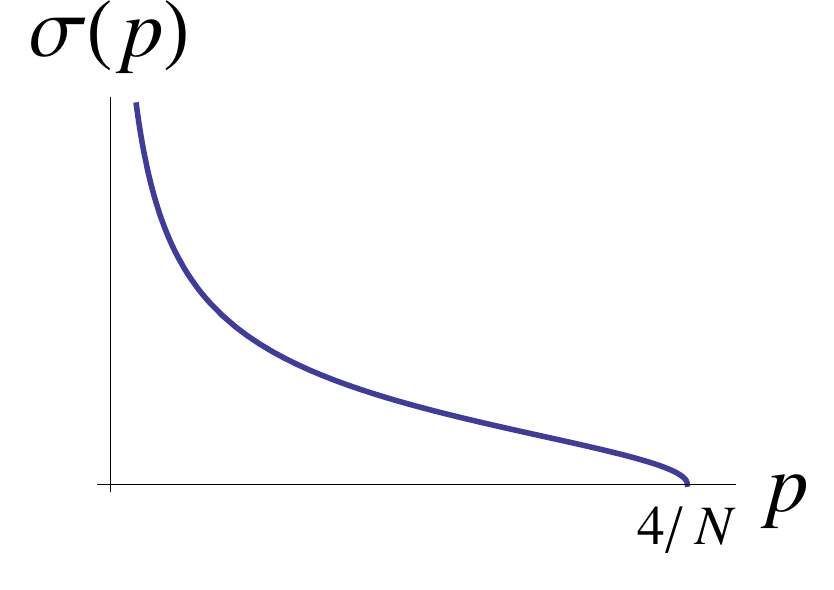}
\includegraphics[width=0.3\textwidth]{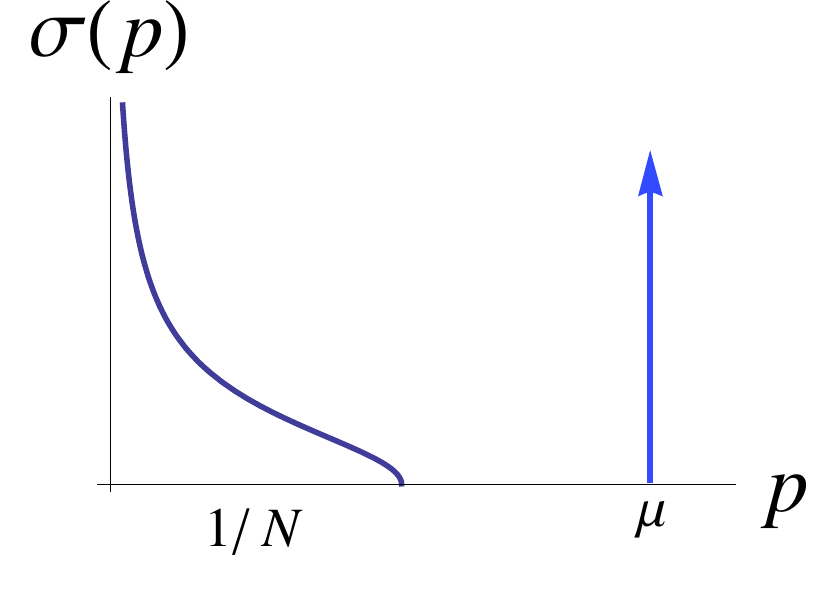}
\caption{
Qualitative entanglement spectra at $N_A=N_B=N$ for
states in the (from left to right): 
phase containing maximally entangled states; 
phase containing typical states;
phase containing separable states (not in scale).}
\label{fig:phases}
\end{figure}

As already mentioned, the entanglement spectrum has been characterized as a function of the purity \cite{G07,Fetal08},
the von Neumann entropy \cite{Florio}, and all Renyi entropies with $q > 1$ \cite{NMV10}
(see also \cite{Fabio}). 
It has been observed that the entanglement spectrum 
changes abruptly in correspondence with two {\it critical values} of the Renyi entropy.
This property defines three {\it phases} with different entanglement features \cite{Fetal08,NMV10,DeP,NMV11}.
The phase corresponding to large values of the entropy contains maximally entangled states.
The central phase contains the typical states.
Finally, the third phase contains separable states.
In particular, while in the first two phases all the squared Schmidt coefficients are typically $\Omega(1/N_A)$,
for low values of the entropy there is a finite probability that a single Schmidt coefficient equals $\mu = \Omega(1)$.
(The latter phase contains a rich structure of metastable configurations corresponding
to local minima of the energy, see \cite{DeP}.)
The qualitative features of the entanglement spectrum in the three phases are
depicted in Figure \ref{fig:phases}.

\section{Typical entanglement as an approach to thermodynamics}\label{thermosec}

\begin{quote}
{\em This fundamental law [that systems are in a Gibbs thermal state] is the summit of statistical mechanics, and the entire 
subject is either the slide-down from this summit, as the principle is applied to 
various cases, or the climb-up to where the fundamental law is derived and the 
concepts of thermal equilibrium and temperature T clarified.} \\
\hspace*{\stretch{1}} Feynman, Statistical Physics, Benjamin-Cummings, 1972
\end{quote}
The phenomenon of typical entanglement has been advocated as an alternative to approaches based on ergodicity and mixing to ascend Feynman's summit by probing and justifying the thermal state assumption \cite{Lloyd88, L78, mahler, GOM01, PSW06}.

To understand whether the thermal state can be justified in terms of typical entanglement we begin by considering the cleanest case wherein all states of the system have the same energy (such that the Hamiltonian is proportional to the identity, e.g. $H=0$.) In this case the Gibbs thermal state $\rho_{th}(\beta, H)=\sum_i \frac{\exp(-\beta E_i)}{Z}
\ket{i}\bra{i}$ reduces to $\rho_{th}(\beta, H)=\sum_{i=1}^{N_A} \frac{1}{N_A}\ket{i}\bra{i}=\frac{I}{N_A}$, where $N_A$ is the dimension of the system in question. For our purposes it is crucial to note that this is equivalent to the reduced state ${\tr}_B \rho_{AB}$ in the event that $\rho_{AB}$ is pure and $A$ and $B$ are maximally entangled (assuming $N_B \geq N_A$), since pure states are defined to be maximally entangled when the von Neumann entropy of the smaller subsystem is maximal and this is the case precisely for the state of the smaller system being $\rho_A=\frac{I}{N_A}$. Thus --in this case of $H=0$--an argument that says that the subsystem should be maximally entangled with the environment would also imply that the system is in a thermal state. One can then use the typical entanglement phenomenon --for dimensions of A and B where it exists-- to say that typically the smaller system should be in a thermal state. 

In the case of a non-trivial Hamiltonian, the argument for justifying the Gibbs state in terms of typical entanglement is less direct and still a topic of active research. In Refs.\cite{Lloyd88, L78, mahler, GOM01, PSW06} the following line is taken. The principle of equal a priori probability states that all allowed states, given e.g.\ a restriction on the energy of the total system ($A$) + environment ($B$) state, are equally likely, i.e. that $\rho_{AB}=\frac{I}{N_R}$ where $N_R$ is the dimension (the minimal number of basis vectors) needed to express the most general state satisfying the restriction (e.g.\ $H\ket{\psi}_{AB}=E\ket{\psi}_{AB}$). Under certain additional assumptions it is well-established that the principle of equal a priori probability implies the Gibbs state on the smaller system $A$. Now, they show, if one rather than relying on the principle of equal a priori probability considered states $\ket{\psi}_{AB}$ from the Haar measure, then the reduced states are typically the same, i.e.\ typically ${\tr}_B\frac{I}{N_R}\approx Tr_B\ket{\psi}_{AB}\bra{\psi}$. 
To gain an intuition for why this is the case, note firstly that for some given observable, if we pick a large typical state it is very unlikely that the observable carries predictability. On that particular measurement it is thus very unlikely that there is a difference to the uniform distribution on the restricted state space. 
Secondly, recall again that the number of local measurements on $A$, i.e.\ of the form $g_A\otimes I$ is very small and fixed by the subsystem size, so that increasing the environment system makes it increasingly unlikely that any of these will carry predictability. If none of them carry predictability the state is indistinguishable from the reduced state of the uniform distribution. 

For completeness, even though this argument is independent of typical entanglement, we briefly give a standard argument for how to get the canonical Gibbs state once one has the uniform distribution over states with a given energy (or in a given energy interval), see e.g.~\cite{Lloyd88}. Firstly assume that the interaction part of the Hamiltonian is so weak that the energy eigenstates of energy $E$ can be well approximated as $\ket{E_i}\ket{E-E_i}$. There may, crucially, be degeneracies; we assume for simplicity that these are only on the environment system (it can easily be generalised), so that the states are labelled $\ket{E_i}\ket{E-E_i,j}$, with $j=1,2,..j_{max}(i)$ and the total dimension of the subspace is $\sum_i j_{max}(i)$.  Then 
\begin{equation}
{\tr}_E (I/d_E)=\frac{1}{\sum_i {j_{max}(i)}}\sum_i j_{max}(i) \ket{E_i}\bra{E_i}.
\end{equation}
Now make the second crucial assumption that  $j_{max}(i)$ is large and only changed slightly by altering i, such that  $j_{max}(i):=j_{max}(E-E_i)\approx j_{max}(E)-E_i j_{max}'(E)\approx j_{max}(E)\exp (-E_i \frac{j_{max}'(E)}{j_{max}(E)})$. Now {\em define} the inverse temperature as $\beta:=\frac{j_{max}'(E)}{j_{max}(E)}$. The Gibbs state follows.

One can identify certain advantages and disadvantages with justifying the thermal state in this manner.  
If one models a quantum system interacting with an environment it is natural to expect them to become entangled so the idea that thermalisation is associated with entanglement with the environment seems to follow naturally from assuming that quantum theory is a good model of reality. Moreover it gives thermalisation more of an objective character than e.g.\ Jayne's maximum entropy principle~\cite{Jaynes57}. This is because if two systems are highly entangled the outcomes of measurements on the individual systems cannot be predictable to any external observer, whereas Jaynes' argument is concerned with classical subjective lack of knowledge. At the same time there is an arbitrariness regarding the measure from which to pick the global pure state. The Haar measure is mathematically natural but there are several physical issues. Many of these have already been mentioned in this review, and it seems the conclusions often survive adding further physical restrictions to the set of states, but understanding how physical the typical entanglement approach is remains an area of active research.  

A key original paper on typical entanglement concerned black hole thermodynamics, more specifically black hole formation and  evaporation \cite{P93bh}; very similar ideas appeared already in~\cite{Lloyd88}. A starting point was to treat a black hole formation and evaporation process as fully quantum and in particular assume that the total formation and evaporation process acts as a unitary evolution on the systems involved. This is in contrast to Hawking's original semi-classical description where the total evolution is modelled as irreversible. The basic idea of Page's model for black hole formation and evaporation is that there is some matter sitting in space which collapses to a black hole, undergoes a unitary evolution, and then gradually splits into an increasing number of subsystems outside the hole and a decreasing number inside the hole. At the level of the quantum model, the total system is a pure state $\ket{\psi}^{(0)}$ which undergoes a unitary evolution to $\ket{\psi}^{(1)}=U\ket{\psi}^{(0)}$. Page models this $U$ as picked from the Haar measure. Next we introduce a partition of the system into two parts, representing the inside and outside of the black hole event horizon as described by a distant observer. This partition is gradually shifted to represent subsystems leaking out of the hole. Now the typical phenomenon tells us that the entropy of the subsystem outside will typically increase until the dimension of the outside equals the inside (this is sometimes called the Page-time), and then decrease towards 0. This means that any information encoded in the choice of initial pure state of the total system is recovered in principle after the evaporation is complete, according to this model. At the same time, as Page points out, if one only looked at the reduced states of the subsystems as they come out it would appear all information is lost, as it is hidden in the correlations. Thus, he argues, calculations which show entropy increase in the reduced states of subsystems do not prove that information is lost. In fact typically states have both maximally mixed subsystems {\em and} maximal entanglement in which the information about the original state is encoded. 
Apart from those early insights random unitaries and typical entanglement remain an important part of the discourse concerning the Black Hole information paradox, see for example\cite{BraunsteinP13}, which states
\begin{quote}
\emph{Our best understanding of the growth and decay of the (von Neumann) entropy of
black hole radiation comes from models based on random unitaries.}
\end{quote}

There is in addition a more general perspective on typical entanglement which is finding use both in thermodynamics and black hole arguments, going by the name of the decoupling theorem, see e.g.~\cite{HaydenHWY08}. The standard description above concerns a bipartite state $\rho_{AB}$ as described by some implicit observer $C$. One may view the decoupling theorem as concerning the same scenario but viewed from a second external observer who also includes $C$ in the description. When A and B are maximally entangled, so that $C$ assigns maximal entropy to $A$, this would from the external observer's perspective be described as the state on $AC$ being 
\be
\rho_{AC}=I_A\otimes \rho_C.
\ee
One says that $A$ is {\em decoupled} from $C$. The decoupling theorem is a proven quantification of the statement that when a Haar random unitary is applied to $AB$, $A$ typically becomes decoupled from $C$. More specifically, 
$\rho_{AC}\approx \frac{I}{N_A}\otimes \rho_C$, provided that the dimension $N_{B}$ satisfies
\be
2\log{N_{B}}\gg\log{N_A}+\log{N_C}-\log\tr\left((\rho^{AC})^2\right).
\ee

This scenario is more general than the standard typical entanglement scenario in that there may be entanglement between $AB$ and $C$, whereas writing down $\rho_{AB}$ as a well defined state as assigned by $C$ implicitly assumes this is not the case (it implicitly assumes the $ABC$ state is `quantum-classical' so that the conditional reduced states on $AB$ corresponding to different states of $C$ are well-defined.). The decoupling theorem is for example used to make a more sophisticated version of Page's argument~\cite{HaydenP07}. In this scenario, the black hole has formed and been evaporating for some time in the same manner as in Page's argument, but then a new system is introduced, which consists of two maximally entangled halves (one representing some secret information and the other half the record thereof). One of the entangled halves is thrown into the black hole and one half kept outside as a reference 
($R$) and the question is how quickly, in terms of number of subsystems emitted, this information comes back out of the black hole, meaning how quickly $R$ is purified by systems outside of the black hole. Haar unitary black hole evolution is again applied to the new larger black hole. The decoupling theorem can then be used to show that this information may leak out very quickly as a function of the number of subsystems emitted, hence the expression `black holes as mirrors'. To see this more concretely, let $A$ be the remnant of the Black hole and $B$ the part emitted (since after the diary was thrown in). The decoupling theorem then says that for sufficiently large $B$, $A$ will be decoupled from $R$, i.e. $\rho_{A,R}\approx \rho_{A}\otimes \rho_{R}$. This means that $R$ must be purified by something outside of $A$: one possible purification of $\rho_{A}\otimes \rho_{R}$ is a product state $\ket{\psi}_{A T}\otimes \ket{R,T'}$ where $T$ and $T'$ are not overlapping with $A$ and thus outside the black hole. For this purification $R$ is indeed purified by something outside of $A$ and all purifications are equivalent up to a unitary on the purifying system ($TT'$), which cannot change the entanglement between $A$ and $R$. We also note that another version of the decoupling theorem is formulated in~\cite{Braunstein13} to describe the evaporation of black holes with trans–event horizon entanglement and provide a potential solution to the black–hole firewall paradox~\cite{Braunstein13,Almheiri13}.

To end this section we remark that black holes are one of the few scenarios where one might not want to assume that quantum theory holds, and two papers consider the same kind of question for probabilistic theories with reversible dynamics more generally, in what is known as the convex framework \cite{MDV12,MOD12}. Generalisations of the typical entanglement statement \cite{MDV12} and the decoupling theorem \cite{MOD12} are proven in those theorems, elucidating which features of quantum theory are involved in this phenomenon. In the case of black holes it is for example shown that one may imagine a non-quantum world inside the black hole which holds on to most of the information even after most subsystems have leaked; basically this could happen if the subsystems inside the hole have more free parameters than in the quantum case \cite{MOD12}.


\section{Random Quantum States for Continuous Variable Systems}\label{cvsec}
 
The starting point for the discussion of entanglement typicality in $N$
dimensional quantum system is the probability density distribution of the (squared)
Schmidt coefficients across a bipartition of the system. 
One natural way to think at the $N$ dimensional system is as a collection of
$\log_2{N}$ qubits, that are then split into two subsets.

A different approach has to be taken if instead one wants to describes a 
system composed of $n$ CV subsystems (introduced in Section \ref{CVsubsec}).
A case of special interest is that of  Gaussian states for the $n$ CV quantum subsystems.
In such case, given a bipartition of the system in $n_A$ and $n_B$ CV systems
(with $n_A + n_B = n$ and $n_A \leq n_B$), the relevant object is the probability
density distribution of the symplectic eigenvalues of the subsystem $A$.
In order to compute it one has to recall that Gaussian states are the submanifold of
states that are obtained by applying the subgroup of Gaussian unitaries on the vacuum state, that is,
\be
\ket{\Psi}_G = U_G \ket{0} \, .
\ee
Let us also recall that Gaussian unitaries are of the form  \cite{B05}
\ba\label{UGhom}
U_G &=&\exp{\left( -i \sum_{i,j=1}^n Y_{ij} a_i^\dag a_j \right)} \exp{\left( \sum_{k=1}^n s_k a_k^2 - s_k (a_k^\dag)^2 \right)}\nonumber\\
&\times& \exp{\left( -i \sum_{i,j=1}^n Y'_{ij} a_i^\dag a_j \right)} \, ,
\ea
where $a_i, a_i^\dag$s are mode ladder operators, $Y$ and $Y'$ are Hermitian matrices, and $s_1, s_2, \dots, s_n$ are real and nonnegative parameters. 
The Gaussian unitaries of the form $\exp{\left( -i \sum_{i,j=1}^n Y_{ij} a_i^\dag a_j \right)}$,
with $Y$ Hermitian, define a representation of the group $\mathrm{U}(n)$. Eq.~(\ref{UGhom}) 
expresses the well known Euler, or Bloch Messiah, decomposition of symplectic operations in 
passive optical elements (essentially beam splitters and phase plates in the quantum optics laboratory, parametrised here 
by $Y$ and $Y'$) and single-mode squeezing operations (implemented, optically, in parametric down conversion 
processes through nonlinear crystals, and represented here by the parameters $\{s_k\}$).

The invariance of Gaussian states under the action of Gaussian unitaries allows one to
compute the probability density distribution of the symplectic eigenvalues $\nu_1, \nu_2, \dots, \nu_{n_A}$
on the manifold of $n_A + n_B$ Gaussian states ($n_A \leq n_B$).
One then obtains \cite{Letal12}
\begin{equation}\label{invm}
P(\nu) = Z_{CV}^{-1} 
\prod_{h>k=1}^{n_A} \left( \nu_h^2 - \nu_k^2 \right)^2 \prod_{j=1}^{n_A} \nu_j^2 (\nu_j^2-1)^{n_B-n_A} \, d\nu \, ,
\end{equation}
where $Z_{CV}^{-1}$ is the normalization function.
Notice that, due to the fact that the manifold of Gaussian states is unbounded, 
the density $P(\nu)$ cannot be globally normalized. 

To cope with this problem Refs. \cite{SDP07,Setal07} resorted, in line with the strand of works that relates typicality to thermodynamical behaviour 
reviewed in Sec.~\ref{thermosec}, to imposing a statistical (canonical) principle on the desired measure \cite{GOM01,PSW06}:
\begin{quote}
\emph{Given a sufficiently small subsystem of the universe, almost every pure state of the universe is such that the subsystem is approximately in the 'canonical state' $\rho_c$.} 
\end{quote}
The �canonical state� is  the local reduction of the global state picked from a distribution of states with maximal entropy under the constraint of expectation value for the total energy operator $H_G$ given by $E$. This amounts to take a thermal canonical state, namely a Gaussian state with null first moments and covariance matrix $\sigma_c = (1 + T/2)I$.
Here the temperature $T$ is defined by passage to 
the �thermodynamical limit�, that is for $n \to\infty $ and $E \to \infty $, $(E - 2n)/n \to T$ (assuming $k_B =1$ for the Boltzmann constant).  

Then, in analogy with thermodynamics, two privileged options can be considered: introducing a temperature or fixing the energy. These two possibilities correspond to canonical and micro-canonical approaches respectively.

Within the former approach the modes' energies $\vec{E}=(E_1,\ldots,E_n)$ -- defined as in Eq.~(\ref{Vener}) and given, in terms of the variables $s_k$ of Eq.~(\ref{UGhom}), by $E_k = \left({\rm e}^{s_k}+{\rm e}^{-s_k}\right)$ -- are assumed to be distributed accordingly to a probability distribution
\be
dP_c(\vec{E}) = \frac{e^{-(|E|-2n)/T}}{T^n} d\vec{E} = \prod_{j=1}^n\left( \frac{e^{-(E_j-2n)/T}}{T} dE_j\right).
\ee
Actually, this distribution maximises the entropy on the knowledge of the continuous variables $E_j$'s for given average total energy $E_{av}=nT$. 
It follows that the �canonical� average $Q_c(T)$ over pure Gaussian states at temperature $T$ of a quantity $Q(E, U,U')$ determined by the second moments alone will read
\be
Q_c(T ) =\int  d\mu_H (U) \int d\mu_H(U') \int
\frac{e^{-(|E|-2n)/T}}{T^n} dE \, Q(E, U,U') , 
\ee
where the integration over the energies is understood to be carried out over the whole allowed domain.

Focusing on the behaviour of the entanglement of a subsystem of $m$ modes (as quantified by the von Neumann entropy of the reduction describing such a subsystem), keeping $m$ fixed and letting the total number of modes $n \to \infty$, it has been determined the average asymptotic entanglement and rigorously proved that the variance of the entanglement tends to zero in the thermodynamical limit \cite{Setal07}. 

The micro-canonical approach consist in assuming a Lebesgue (`flat') measure for the energies $E_j$'s inside the region $\Gamma_{E}=\{\vec{E} : \sum_{j}E_j\le E\}$. 
Denoting by ${\rm d}P_{mc}(\vec{E})$ the probability of the occurrence of the energies $\vec{E}$, 
one has 
\ba
{\rm d}P_{mc}(\vec{E}) &=& {\cal N} \,{\rm d}^n\vec{E} \equiv {\cal N}\,{\rm d}E_1\ldots{\rm d}E_n   \quad {\rm if} \quad \vec{E}\in \Gamma_{E}\, , \nonumber \\
{\rm d}P_{mc}(\vec{E}) &=& 0 \quad {\rm otherwise} \; ,
\ea
where ${\cal N}$ is a normalisation constant equal to the inverse of the volume of $\Gamma_{E}$.
Such a flat distribution maximises the entropy in the knowledge of the local energies 
under the constraint of fixed total energy. 

The micro-canonical average $Q_{mc}(E)$ over pure Gaussian states at 
maximal energy $E$ of 
the quantity $Q(\vec{E},\vartheta)$ determined by the second moments alone 
will thus be defined as 
\be
Q_{mc}(E) = {\cal N}\int {\rm d}\mu_{H}(U) \int_{\Gamma_{E}} {\rm d}\vec{E} \, Q(\vec{E},\vartheta) \; ,  \label{micro}
\ee
where the integration over the Haar measure is understood to be carried out over the whole compact domain 
of the variables $Y_{ij}$ of Eq.~(\ref{UGhom}), compactly represented by the unitary $U$ in agreement with the convention above.
The normalisation can be easily determined as ${\cal N}=n!/(E-2n)^{n}$ and 
leads to a marginal density of probability $P_{n}(E_j,E)$ for each of the energies $E_j$ given by
\be
P_{n} (E_j,E) = \frac{n}{E-2n} \left( 1- \frac{E_j-2}{E-2n} \right)^{n-1} \; . \label{marginal}
\ee
Although the energies $E_{j}$ are not independently and identically distributed for finite $n$, 
they are so in the thermodynamical limit, where one has $P_{n} (E_j,E)\simeq {\rm e}^{-\frac{E_j-2}{T}}/T$ 
by defining $T=(E-2n)/n$. 
The equivalence of statistical ensembles of classical thermodynamics is thus recovered, and the general canonical principle fulfilled.

The average purity of a subsystem of $n$ modes may be determined exactly under both the canonical and the micro-canonical measures. Considering, for simplicity, a single mode subsystem of an $n$-mode system, and denoting its canonical and micro-canonical average purity by, respectively ${\cal P}_{c}$ and ${\cal P}_{mc}$, one has \cite{Setal07}
\be
{\cal P}^{-2}_{c} = \frac14 \frac{n-1}{n+1}(T^2+4T) + 1  \, ,
\ee
where $T$ is the temperature associated with the canonical measure, and
\be
{\cal P}^{-2}_{mc} = 
\frac{(n-1)}{4(n+2)(n+1)^2}\Big(\tilde{E}^2+4(n+2)\tilde{E}\Big) + 1 \, ,
\ee
where $E$ is the micro-canonical energy. As a reference value, the maximal purity ${\cal P}_M$ for 
given energy $E=\tilde{E}+2n$ obeys the relationship
\be
{\cal P}^{-2}_M = \frac{(\tilde{E}+4)^2}{16} \; . \label{minpur}
\ee
The standard deviations on such average quantities may also be derived analytically, 
by evaluating the fourth order moments:
\ba
{{\cal P}^{-4}_{c}} = \frac{1}{16}\frac{n! (n-1)}{(n+3)!}
\Bigg[(n^2+11n+22) T^4 +8(n^2+8n+6) T^3 +8(3n^2+15n+10) T^2 \nonumber\\
 +32 (n+3)(n+2) T \Bigg] +1 .
\ea
and
\ba
{\cal P}^{-4}_{mc} &=& 
\frac{(n!)^2(n-1)}{16(n+4)!(n+3)!}\left( (n^2+11n+22)\tilde{E}^4 +8(n+6)(n+4)(n+1)\tilde{E}^3 
\right. \nonumber \\
&&\left. + 8(n+4)(n+3)(3n^2+15n+10)\tilde{E}^2 +32(n+4)(n+3)^2(n+2)^2\tilde{E} \right) + 1 .
\ea

The micro-canonical mean ${\cal P}_{mc}^{-2}$ is increases monotonically with $E$ for fixed $n$ and, 
for $n>2$, decreases monotonically with $n$ for given $E$  
(although a peculiar finite size effect is apparent for 
$\tilde{E}\le10$, with ${\cal P}_{mc}^{-2}$ that increases in going from $2$ to $3$ modes).
This general trend just reflects the fact that more available energy generally allows for higher entanglement, 
while the presence of more modes drains energy away to establish correlations which do not involve the 
particular chosen mode.
The canonical average entanglement is instead monotonically increasing 
with both temperature and number of modes.
This behaviour is encountered also for the micro-canonical entanglement 
with given maximal total energy {\em per mode}, which is, even for small $n$, 
very similar to the canonical average (upon replacing $\tilde{E}/n$ with $T$).
The increase of the average canonical entanglement with increasing number of modes but 
fixed temperature is a non trivial, purely `geometric' effect, 
due to the average over the Haar distributed compact variables. 
An analogous increase is observed assuming a given, fixed value for the squeezing variables 
and averaging only over the compact variables: as the number of total modes increases, 
a given mode has more possibilities of getting entangled, even keeping a fixed mean energy per mode.
The standard deviations above are generally increasing with total energy and temperature for fixed total number of modes 
(as more energy allows for a broader range of entanglement). 
Significantly, these partial analytical results clearly show the arising of the concentration of measure 
around a thermal average. Both for the canonical case and for the micro-canonical one with 
$\tilde{E}=nT$ the standard deviation decreases with increasing number of modes $n$, 
falling to zero asymptotically
(after transient finite size effects, for very small $n$). 

Moreover, in the micro-canonical instance, the thermal average 
of concentration is generally very distant, even for relatively small $n$, from the allowed maximum 
of (\ref{minpur}) (which clearly diverges in the thermodynamical limit): {\em e.g.}, 
for $\tilde{E}=10n$ one has that the average ${\cal P}_{mc}^{-2}$ is, respectively $16.5$  
and $257.1$ standard deviations away from the maximal value ${\cal P}^{-2}_{M}$ for
$n=5$ and $n=20$. Such a distance increases monotonically with the total number of modes.
This highly peaked concentration for finite $n$ allows one to  
obtain strict upper and lower bounds on the average von Neumann entropy of entanglement of single-mode 
subsystems \cite{Setal07}.

 
\section{Outlook}\label{conclusec}

This article is meant to be a summary of mathematical findings related to the notion of typicality of composite quantum states. 
This approach to the study of local entropies has produced an established framework that can now be applied in several fields
of mathematical and fundamental physics. The phase transitions discussed in Section \ref{statsec} are purely classical, but quantum 
entanglement is liable to provide one with a signature for {\em quantum} phase transitions \cite{ON02}: investigating possible links between typicality and quantum state transitions would hence be worthwhile. In a different direction, a phase transition occurring for typical purity \cite{Fetal08} can be paralleled to a well known phase transition occurring in conformal field theory \cite{DGZ95}. As already mentioned in Section \ref{thermosec}, the typicality approach to thermodynamics could be extended to the notion of black hole entropy. The notion of typical entanglement for continuous variable systems should also be developed further, and more explicitly connected to random processes and thermodynamics.


\appendix

\section{Random mixed states}\label{appendix}

In this appendix, we sketch a couple of possible approaches to the issue of mixed random states of finite
dimensional systems (for a more general approach to the topic, please refer to \cite{Zyco};
see also \cite{DeP2} for the local purity distribution of globally mixed random states).
A way to define them is to borrow results from Sec.\ref{randsec} about the induced measure from random pure states.
Specifically, suppose to write a mixed state $\rho$ of a system $S$ as
coming from a pure state $|\psi\rangle_{SE}\in 
\mathcal{H}_S \otimes \mathcal{H}_E$ by tracing away the environment $E$,
\be
\rho=\tr_E |\psi\rangle_{SE}\langle\psi|,
\ee
where $|\psi\rangle_{SE}$  is picked up 
randomly according to the measure (\ref{dmuCP}). As seen in Sec.\ref{randsec}, this induces a measure on the density operators of the system $S$. Let us now split the latter into two further subsystems, considering 
${\mathcal H}_S = \mathcal{H}_A \otimes \mathcal{H}_B$. Then we can look for entanglement 
 of the random induced states on ${\mathcal H}_S$ with respect to the $A-B$ bipartition.
In Refs.\cite{ASY12,ASY14}, using tools from high-dimensional convexity 
(i.e. asymptotic geometric analysis),
 it has been shown that random induced states on ${\mathcal H}_S= \mathcal{H}_A \otimes \mathcal{H}_B$ exhibit a phase transition phenomenon with respect to the dimension $N_E$ of the environment  space. 
Assuming $N_A\le N_B$, one can define a threshold given by $N_A^2N_B$ (up to a poly-log factor):
that is, if one has a system $AB$ comprising a number $n=n_A+n_B$ of individual qubit systems, 
then the two subsystems $A$ and $B$ typically share entanglement if $n_E<2 n_A+n_B$, 
and typically do not share entanglement if $n_E>2n_A+n_B$.  
This approach provides a relationship between entanglement properties and the environment.


Alternatively, to define random mixed states, one can get rid of the environment and 
start from a global mixed state $\rho$ with fixed purity. 
Then, following \cite{MDV12}, let us consider the ensemble of states obtained 
by applying a Haar-distributed unitary, according to (\ref{dmuU}), to $\rho$. 
Specifically, suppose to have a system $AB$ comprising a number $n=n_A+n_B$ of individual qubit systems. 
Any state $\rho$ of such a system may always be expressed in the basis $\{g_j\}$ of the space of Hermitian operators as
\be
\rho = \sum_{j=0}^{2^{2n}-1} \xi_j g_j,
\ee
where each $g_j$ is a tensor product of Pauli operators (and identities).
The operator basis chosen is clearly orthogonal with respect to the Hilbert-Schmidt inner product:
\be
{\rm Tr} (g_j g_k) = 2^{n} \delta_{jk} \quad \forall j,k ,
\ee
and such that $g_0=I_{2^n}$. Then, $\xi_0=\frac{1}{2^{n}}$ by normalisation of the quantum state.
Besides, this basis features the remarkable property that any two $g_j$ and $g_k$ 
for positive $j$ and $k$ are related by a unitary, {\em i.e.} $\exists V\in {\rm U}(2^n):$ $g_j=Vg_k V^{\dag}$, $\forall j,k\ge 1$.

Let us now define the state $\rho$ with a given Renyi-2 entropy: $\tr\rho^2={\cal P}$, and consider the ensemble of states obtained 
by applying a Haar-distributed unitary to $\rho$. Averages with respect to such an ensemble will be indicated by 
$\mathbb{E}$.
We want to determine the average local purity of $n_A$ degrees of freedom under such a global average, at fixed global purity.

A proper ordering of the basis elements allows one to write the local state $\rho_A$, without loss of generality, as:
\be
\rho_A = \tr_B \rho = 2^{n_B} \sum_{j=0}^{4^{n_A}-1} \xi_j g_j,
\ee
(only operators which reduce to the identity on the space of the $n_B$ qubits contribute to the partial trace).
Also, the orthogonality of the basis leads to
\be
\tr\rho_A^2 = 2^{2n_B}2^{n_A}\sum_{j=0}^{4^{n_A}-1} \xi_j^2, 
\label{mua1}
\ee
while the constraint on the global purity implies
\be
{\cal P} = \tr\rho^2 = 2^{n} \xi_0^2 + 2^n \sum_{j=1}^{4^n-1} \xi_j^2 = \frac{1}{2^n} + 2^n \sum_{j=1}^{4^n-1} \xi_j^2. \label{mu1}
\ee
Now, because the basis elements $g_j$ are all unitarily equivalent for $j\ge1$, one has that the Haar averages 
of the coefficients $\xi_j$ must be the same: $
{\mathbb E} \xi^2_j = {\mathbb E} \xi_k^2 
\quad \forall j,k\ge 1.$
Combining this fact with the unitarily invariant constraint (\ref{mu1}) yields
\be
{\cal P} = \frac{1}{2^n} + 2^n(4^n-1) {\mathbb E} \xi_j^2 \quad \Rightarrow \quad {\mathbb E} \xi_j^2 = 
\frac{2^n{\cal P} -1}{2^{2n}(4^n-1)}. 
\ee
We can now determine the exact average of the local purity of equation (\ref{mua1}) as 
\be
{\mathbb E} \tr\rho_{A}^2 = \frac{1}{2^{n_A}} + 2^{n_B}({\cal P} -1/2^n) \frac{4^{n_A}-1}{(4^n-1)},  \label{mua2}
\ee
{which may be recast as  $\frac{{\mathbb E} \tr\rho_{A}^2 - 2^{-n_A}}{{\cal P} - 2^{-n}} = 2^{n_B} \frac{4^{n_A}-1}{4^{n}-1}.$ 
This shows that the ratio between the average deviation from the minimal possible purity of a subsystem and the deviation from the 
minimum possible global purity is proportional to the ratio between the number of local and global degrees of freedom.
It is also apparent that, regardless of the global purity $\cal P$ and in agreement with asymptotic considerations, the local purity tends to its minimum value
in the limit where $n_B$ goes to infinity and $n_A$ stays finite. Even in the asymptotic case where $n$ goes to infinity but the ratio between $n_A$ 
and $n$ tends to a constant, the local purity tends to the minimum value (zero, in this instance).
In the case of a global pure state (${\cal P}=1$), Eq.~(\ref{mua2}) simplifies to (\ref{Pave}).

\acknowledgments
OD acknowledges support from EU Network TherMiQ.
CL thanks F. Delan Cunden for useful comments.

\end{document}